\documentclass[12pt,a4paper]{article}
\usepackage{cmap}
\usepackage[T2A]{fontenc}
\usepackage[cp1251]{inputenc}
\usepackage[english, russian]{babel}
\usepackage{amsfonts,amsmath,amssymb,amsthm,longtable,hhline}
\usepackage{bm}
\usepackage{cite}
\usepackage{color,soul}
\usepackage{color,xcolor}
\usepackage{multicol}
\usepackage{tabularx,caption,multirow}
\usepackage{graphicx}

\usepackage[labelformat=empty,labelsep = none,skip=0pt,justification=raggedright,singlelinecheck=false]{caption}

\usepackage[linktoc=none,linktocpage=true,unicode,bookmarks,bookmarksopen,bookmarksopenlevel=2,colorlinks,linkcolor=blue,citecolor=blue,urlcolor=blue]{hyperref}

\makeatletter
\renewcommand{\@biblabel}[1]{#1.} 
\makeatother
\theoremstyle{remark}

\newcounter{urav}[section]
\newcounter{resh}[urav]

\newcounter{exmp}

\let\ds=\displaystyle

\def\fracskip{\mskip 1mu \relax}
\def\nfrac#1#2{{\fracskip#1\fracskip\over\fracskip#2\fracskip}}
\def\dfrac#1#2{{\ds\nfrac{#1}{#2}}}

\let\frac=\nfrac
\def\pd#1#2{\dfrac{\partial#1}{\partial#2}}
\def\pdd#1#2#3{\ifx#2#3\pd{^2#1}{#2^2}\else\pd{^2#1}{#2\partial#3}\fi }

\newcommand{\clh}[1]{\colorbox{yellow}{#1}}%
\newcommand{\clhp}[1]{\colorbox{yellow}{\parbox{\textwidth}{#1}}}%
\newcommand{\red}[1]{\textcolor{red}{#1}}

\begin{document}
\large 

\centerline{\bf\Large  The similarity index of scientific publications} 
\centerline{\bf\Large  with equations and formulas, identification of self-plagiarism,}
\centerline{\bf\Large  and testing of the iThenticate system\clh{$^*$}}

\bigskip

\centerline{Andrei D. Polyanin$^{b}$, Inna K. Shingareva$^{b}$}
\medskip
\centerline{\it $^a$ Ishlinsky Institute for Problems in Mechanics RAS,}  
\centerline{\it 101 Vernadsky Avenue, bldg 1, 119526 Moscow, Russia}
\centerline{\it $^b$ Department of Mathematics, University of Sonora,} 
\centerline{\it Blvd. Luis Encinas y Rosales S/N, Hermosillo C.P. 83000, Sonora, M\'exico}
\bigskip
\bigskip

\let\thefootnote\relax\footnotetext{
\hskip-20pt\clhp{$^*$ 
This text is a free author’s translation from Russian into English of the article 
\textit{The similarity index of mathematical and other scientific publications with equations 
and formulas and the problem of self–plagiarism identification}
by A.D. Polyanin and I.K. Shingareva, Mathematical Modeling and Computational Methods, 2021, No. 2, pp.~96--116
[А.Д. Полянин, И.К Шингарева, \textit{Индекс подобия математических и других научных публикаций 
с уравнениями и формулами и проблема идентификации самоплагиата},
Математическое моделирование и численные методы, 2021, № 2, с. 96–116; https://mmcm.bmstu.ru/articles/253].}}

The problems of estimating the similarity index of mathematical and other scientific publications
containing equations and formulas are discussed for the first time.
It is shown that the presence of equations and formulas (as well as figures, drawings, and tables)
is a complicating factor that significantly complicates
the study of such texts. It is shown that the method for determining
the similarity index of publications, based on taking into account individual
mathematical symbols and parts of equations and formulas, is ineffective and
can lead to erroneous and even completely absurd conclusions. The possibilities
of the most popular software system iThenticate, currently used in scientific journals,
are investigated for detecting plagiarism and self-plagiarism.
The results of processing by the iThenticate system of specific examples and
special test problems containing equations (PDEs and ODEs), exact solutions, and some formulas are presented.
It has been established that this software system when analyzing inhomogeneous texts,
is often unable to distinguish self-plagiarism from pseudo-self-plagiarism (false self-plagiarism).
A model complex situation is considered, in which the identification of self-plagiarism
requires the involvement of highly qualified specialists of a narrow profile.
Various ways to improve the work of software systems for
comparing inhomogeneous texts are proposed.
This article will be useful to researchers and university teachers in
mathematics, physics, and engineering sciences, programmers dealing with problems in image
recognition and research topics of digital image processing, as well as
a wide range of readers who are interested in issues of plagiarism and self-plagiarism.
\medskip


\section{Introduction}

\subsection{Preliminary remarks and some concepts}

This article will focus on research papers and books in various areas of
mathematics and natural sciences, which contain a significant number of
equations and formulas.
The presence of equations and formulas is a factor that significantly
complicates the estimation of the volume of borrowings and the similarity index in such publications.

It is important to note that the estimation of the similarity index
and the identification of self-plagiarism of publications with equations and formulas
is a relevant, complex, and very delicate topic
that is practically not covered in the literature and has not been widely discussed by the scientific community.

Let us define some concepts that are often used below
(some of them are introduced for the first time).

\textit{Equation} (in mathematics and physics) is an analytical expression consisting of
letters (usually Latin and Greek alphabets), numbers, mathematical symbols, and operators,
which contains an equal sign
and connects known and unknown (sought) quantities.

\textit{Formula} is a symbolic-analytical expression describing the relationship
between various variables and/or constant values.

\textit{Remark 1.} The terms \textit{equation} and \textit{formula}
are close in meaning and the authors often do not distinguish them. However, when formulating problems
in which an unknown quantity appears, it is preferable to talk about equations.
In this article, we use the terms equations and formulas in a broad sense, by default adding to them also inequalities, identities,
solutions, transformations, differential and functional constraints, boundary and initial conditions,
and any other mathematical constructions that are written in the symbolic-analytical form.

\textit{Homogeneous text} is a text consisting only of separate letters, words, and sentences
(without equations, formulas, graphs, figures, and tables).

\textit{Inhomogeneous text} is a text consisting of separate letters, words, sentences, equations, and formulas
(may also include figures, drawings, and/or tables).

\textit{Plagiarism} is a direct borrowing of parts of the text of articles and books
written by other authors without the necessary references. Paraphrasing a significant fragment
of other authors' works by changing individual words (as well as mathematical symbols and letters
in scientific texts with equations and formulas) and their order
while maintaining the logical structure of the argumentation
is also plagiarism if there are no references to the work that was used.
The characteristic features, types, and ways of identifying plagiarism, as well as related issues are discussed in
\cite{vrb2017,lyk2016,dob2015,adithan2018,gel2020,mem2020,roberts2020}.

\textit{Self-plagiarism} is the reuse by the author of significant, identical, or nearly identical
parts of his own texts from earlier works without reference to the original source.
The reasons for the widespread self-plagiarism and ways to combat it, as well as related issues, are discussed in
\cite{bre2009,kot2011,mar2013,kyl2019,lin2020,iThb}.

\textit{Similarity index} is a value defined as the number of words in the author's text,
which coincides with the number of words in the sources taken for comparison,
referred to the total number of words in the author's text and multiplied by 100\%
(see \cite{adithan2018, uw-turnitin2016}).
This definition is valid only for homogeneous texts in which there are no equations, formulas, and figures.

\textit{Pseudo-plagiarism} is a seeming (false, imaginary) plagiarism.
The author's text is characterized by original ideas and/or new content and
a significant number of coincidences of individual words, small verbal phrases,
individual fragments of equations, and parts of formulas found
in works of other authors. The coincidences are of a technical nature
(play a secondary role) and do not affect the main content of the publication.

\textit{Pseudo-self-plagiarism} is a seeming (false, imaginary) self-plagiarism.
The author's text is characterized by original ideas and/or new content and a significant number
of coincidences of individual words, small verbal phrases, individual
fragments of equations, and parts of formulas found in other works of the same author.
The coincidences are of a secondary, purely technical nature.

\subsection{iThenticate system}

Various scientific journals employ different plagiarism checking systems, for example,
\textit{iThenticate}, \textit{eTBLAST}, \textit{SPlaT}, \textit{CrossCheck},
\textit{Turnitin}, and \textit{WriteCheck}~\cite{lin2020,zha2012} (see also \cite{adithan2018}).

iThenticate is a commercial software system for detecting plagiarism and self-plagiarism,
which is sold to publishers, universities, research institutes, news agencies, corporations, law firms, and others~\cite{iTh}.
Currently, the iThenticate system (or shortly \textit{iTh-system}) is the most popular and
the most powerful system for checking English-language texts\footnote{Although the iTh-system works in 30 languages,
the system can only match texts in the same language~\cite{teixeira2020}.
In general, the existing similarity detection software
is still unable to detect cross-language plagiarism or
self-plagiarism.}, which is actively used in
scientific journals to reject articles with
a high similarity index\footnote{For scientific journals published by \textit{Elsevier} and \textit{Springer}, the maximum allowable
similarity index of an article is usually 15\% (excluding references)~\cite{els}.}
at the initial stage (before a reviewing process).
A brief description of the working of the iTh-system and specific examples
of its application can be found in \cite{gre,rus2017}.

In general, the use of the iTh-system has shown its high efficiency in detecting
plagiarism and self-plagiarism of scientific publications consisting of homogeneous texts
in various fields of humanities and social sciences,
including economic and legal sciences, as well as in medicine and biology.

Note that the Th-system, after processing the article, gives out its text,
in which words and sentences coinciding with the words
and sentences of other articles and books by various authors selected from the corresponding databases,
including many electronic publications, are marked in color.
In the case of self-plagiarism analysis, the text of the article in question
is compared with the texts of other publications of the same authors.

Currently, the editors of many scientific journals, even before reviewing articles,
use the iTh-system to analyze their similarity index and immediately reject articles
with a high similarity index (and often the results of the work of the iTh-system are not sent to authors,
they are only told that the article has a high similarity index).

In mathematics and natural sciences, the use of the iTh-system and other existing software systems to determine
the similarity index of publications with inhomogeneous text can lead to erroneous and
even completely ridiculous conclusions.
An exploratory analysis of different forms of plagiarism (explicit and disguised)
observable in mathematical publications is presented in~\cite{schubotz2019},
where the authors investigated editorial notes from zbMATH and compared them
to the results of current plagiarism detection systems. The investigation of
the selected cases indicates that the application of the text-based iTh-system appears
insufficient for analyzing similarities in mathematical publications.
Some critical comments of scientists regarding the use of the iTh-system
for articles with equations and formulas were also expressed in the ResearchGate scientific network.

We have tested the iThenticate system from different perspectives on a number of
mathematical and physical articles and special test problems containing equations,
formulas, solutions, figures, drawings, and tables.
The results of our analysis and conclusions are summarized sequentially below.

\section{What should be excluded when comparing any texts}

\subsection{It is necessary to exclude from the
comparison everything that is not related to the scientific content of the article}\label{s:terms1}

The names, surnames, working addresses of authors, email addresses,
references to grants and other financial support, acknowledgments,
phrases about the contribution of the authors and the absence of a conflict between them,
as well as any other words and sentences
not related to the scientific content of the article should not be taken
into account when determining the similarity index of the article.

Despite the obviousness of these simple requirements,
such information is now often included when calculating the article similarity index using the iTh-system.

\subsection{It is necessary to exclude scientific terms and stable phrases from comparison}\label{s:terms2}

Scientific terms and stable phrases (a short combination of words) generally accepted in the scientific community
(such as \textit{elementary function, continuous function, scalar product,
ordinary differential equation, Laplace equation, Navier--Stokes equations, Cauchy problem, first boundary value problem,
collocation method, method of matched asymptotic expansions, Fourier method,
existence and uniqueness theorem,  diffusion boundary layer, self-similar solution, Reynolds number,
Euclidean space}, and many others) do not belong to self-plagiarism and should not be taken into account
when determining the similarity index of an article,
since they cannot be replaced with other words without significantly degrading the text.

Unfortunately, at present, scientific terms and stable phrases are not taken into account
by the iTh-system, which often leads to a significant overestimation of the similarity index
of articles and authors can face the undeserved (false) accusation of self-plagiarism
(scientific terms can be removed when comparing texts, but this is a very
painstaking and tedious work that responsible and technical staff
of the journals try to avoid since it has to be done manually).

Moreover, there are no good reasons to consider
simple and often used in mathematics short word expressions and phrases of the type:
\textit{substituting expression (1) into equation (2), we get};
\textit{the solution to problem (3) has the form};
\textit{consider the nonlinear differential equation};
\textit{that which was to be demonstrated};
\textit{where A and B are arbitrary constants};
\textit{transformation of the independent variable};
\textit{it is not difficult to prove that};
\textit{the proof is by direct calculation};
\textit{Figure 1 shows the dependency};
\textit{it is important to note that} etc., as self-plagiarism.

It is important to note that the vast majority of actively published scientists
are gradually developing their own individual author's style of writing texts,
which consists in their more frequent use of specific words, phrases, and short word expressions,
as well as in the choice of methods for constructing sentences and logical structures.
When working on the texts of articles, these authors do not rewrite or copy individual phrases
from their previous publications (namely, this is unreasonably blamed on them
by the developers of the iTh-system), they just write using their own style.
Therefore, the current practice of using the iTh-system is usually
not the identification of self-plagiarism, but an extremely ignorant and
barbaric struggle with the individual style of the authors, which is far
from the same thing.
This approach leads to an artificial restriction of the possibility of individual scientific self-expression and creativity
and can be classified as an obvious violation of elementary authors' rights.
At the same time, the scientific component of publications
fades into the background, and
the low-content purely technical design activity
imposed on the authors, unnecessary for readers and hindering the development of science, becomes the main one.

Let us give a simple analogy that explains the absurdity of the current state of affairs.
Each printed scientific journal has its own individual cover.
Let us now require that the cover design for each issue (or volume)
of the journal is at least 85\% different from the design of the covers of previous issues.
Such a requirement for scientific journals looks wild, but it fully corresponds to
the practice of the iTh-system working with the authors' texts.

Taking into account the above and following~\cite{rus2017}, when using the iTh-system,
it is necessary to exclude short sequences of words (less than eight to ten words in length)
and bibliography from the check. Such options are provided in the iTh-system,
however, it is extremely rare in journals to exclude short sequences of words from consideration
due to the insufficient ability of technical staff to work with this system.
In \cite{rus2017}, it is noted that the exclusion of short sequences of words
and bibliography  can reduce the number of detected matches by one
and a half to two times, even for simple homogeneous texts without formulas.

It is useful to recall that scientists write articles for their colleagues
and interested specialists in related professions, but not for
the developers of the iTh-system.
It is important for readers to understand what new results were obtained by the author of the publication,
while they do not care at all how much individual words and phrases
used in the article differ in form from words and
phrases written by the same author in his previous articles.
The main task of the authors is to get new results and write about
them understandable and clear, and the main goal of the iTh managers
is to obtain maximum profit from collective and individual users of
this system (and business interests here clearly prevail over common sense).

\section{The iThenticate system is unable to adequately compare formulas and equations}

\subsection{Qualitative features of articles with equations and formulas. The problem of determining the similarity index of such publications}

1. In homogeneous publications that contain only words
(but do not contain equations, formulas, figures, drawings, and tables),
the iTh-system determines their similarity index in percentage as
follows\footnote{Here, the simplest method for determining the similarity index is described
(the iTh-system provides for some modifications and complications in calculating the similarity index)~\cite{zhang2014}.}.
The total number of words in the text of the article in question is calculated
that coincide with words in other articles selected from the databases linked to the iTh-system.
Then the total number of such matching words is divided by the total number of words
in the text of this article and the result is multiplied by 100\%.

2. In articles on mathematics and theoretical physics, equations and formulas
usually play a major role and have a greater specific weight than the accompanying verbal
description, which is often secondary and much less important.
Therefore, when researching such publications for self-plagiarism,
equations and formulas should be analyzed first.

As a result, the following important question arises:
how can a single equation or formula be compared with ordinary text
without formulas? For example, the following options are possible:

\begin{description}

\item{(i)} each formula can be considered equivalent to one word;

\item{(ii)} each formula can be considered equivalent to several words;

\item{(iii)} each letter and each mathematical symbol
included in the formula can be considered equivalent to one word;

\item{(iv)} you can replace each formula
with a verbal description, and then count the number of words
in the verbal description.

\end{description}

Each of these options has its own disadvantages.
Option (i) is unsuccessful because it completely depreciates
formulas, which often contain extensive information combining their constituent inhomogeneous parts.
Option (ii) is quite suitable, however, when using it, it is necessary to indicate how many words one formula is equivalent to
(this issue is subject to discussion, see Section~\ref{ss:ways-to-improve} below).
Option (iii) is ambiguous because often the same formula can be written in different ways
(for example, $e^x$ and $\exp{x}$) or represented as several formulas;
other disadvantages of the approach based on comparing individual letters
and parts of equations and formulas are discussed further in Section~\ref{ss:incorrect-work}.
The last option (iv) is ambiguous (the verbal description may be different)
and difficult for practical implementation,
but it correctly reflects the important qualitative difference between the formula and the word: namely,
the formula contains compressed information equivalent to a set of many words.

3. The most important characteristic qualitative feature of articles
with formulas and equations is that one-to-one replacement of all
(or part of) letters in all formulas and equations for any others
(for this purpose, letters of the Latin and Greek alphabets,
and sometimes the Gothic alphabet can be used) does not change
the content of the article.
In other words, two articles that differ only in the notation of letters in formulas
and equations are considered identical.
The iTh-system considers equivalent formulas and equations that differ only in letter swapping to be
different\footnote{Therefore, some authors sometimes use letter swapping
in formulas and equations to decrease the similarity index of articles.}
(and this state of affairs is unlikely to be
significantly changed for the better in the foreseeable future).

This circumstance sharply limits the possibilities of an adequate application
of software systems of the iTh-type for the comparison of
individual formulas in different texts. It is necessary to take the simplest
option as the basis for the operation of such systems with inhomogeneous texts:
two formulas are considered the same if all letters, mathematical symbols and numbers
included in them are the same.
Further in Section~\ref{ss:incorrect-work}, by using specific examples,
it will be shown that even with such a simple way of comparing equations and formulas
in different texts, the iTh-system often makes gross errors,
leading to a significant increase in the similarity index of publications.

\subsection{Examples of incorrect work of the iThenticate system}\label{ss:incorrect-work}

The iTh-system is often unable to adequately compare different,
but similar formulas and equations. This circumstance is currently one,
but far from the only, of the major disadvantages of this system.
Two simple illustrative examples of how the iTh-system misdiagnosed complete plagiarism (identical coincidence)
of various formulas and equations are given below.

\textbf{\textit{Example 1.}}
The iTh-system identifies two different formulas
$$
g=1+|z|+|f|^{1/2}\quad\text{and}\quad g=(1+|z|+|f|)^{1/2}.
$$
For visual evidence of this, see line no. 3 of Test problem~2.

\textbf{\textit{Example 2.}}
The iTh-system also does not distinguish between formulas
$$
y=a+bx^{-1/2}\quad\text{e}\quad y=a+bx-1/2.
$$
For visual evidence of this, see line no. 4 of Test problem~2.

From the above examples, it can be seen that the iTh-system does not know
how to work with parentheses and indices at all
(although even very mediocre school children can do this).
Now let us move on to a more complex example and special test problems.

\textbf{\textit{Example 3.}}
The iTh-system shows that the following two different nonlinear partial differential equations:
\begin{align}
u_{t}=[f(u)u_x]_x+g(u)\quad\text{and}\quad u_{tt}=[f(u)u_x]_x+g(u)
\label{eq:01}
\end{align}
are almost the same because most of their terms are the same.

Here the first equation is a nonlinear reaction-diffusion equation
(parabolic type partial differential equation), and the second equation is
a nonlinear Klein--Gordon type equation describing the propagation of waves
(hyperbolic type partial differential equation); in other words,
these equations are as different from each other roughly like as a cow is from a horse.

We compared two large papers \cite{pol2019a,pol2019b}, containing many numbered equations
and formulas, in which exact solutions of more complicated than \eqref{eq:01}
related nonlinear partial differential equations differing only
in terms of $u_{t}$ and $u_{tt}$  were constructed
(note that these equations do not have the same solutions).
The iTh-system concluded that the articles under consideration
are very similar and have the similarity index of 61\%
(the calculation is based on counting fragments of the inhomogeneous text
of the article \cite{pol2019b} that coincide with the fragments
of the text of the article \cite{pol2019a}).
This ridiculous conclusion is primarily since the iTh-system
compares individual parts of different equations and formulas and considers them
partially identical if at least one term or one letter in them is the same
(then all such pseudo-matches are taken into account when calculating the similarity index).
It is obvious that making general conclusions by comparing
the individual parts of various formulas and equations is entirely absurd.

\textbf{\textit{Test problem 1.}}
In order to more clearly and in more detail demonstrate
the inconsistency of the procedure for determining the similarity index
of inhomogeneous texts using the iTh-system, we now consider
a combined test problem containing two variants of inhomogeneous text,
in which many words are the same, but the equations under consideration
and the solutions obtained are entirely different.
The matching fragments of the compared texts detected by
the iTh-system are highlighted in red (see below).

\begin{figure}[!h]
\noindent
\begin{flushleft}\textit{\normalsize Test problem 1, variant 1.}  \end{flushleft}
\centering
{\includegraphics[width=140mm]{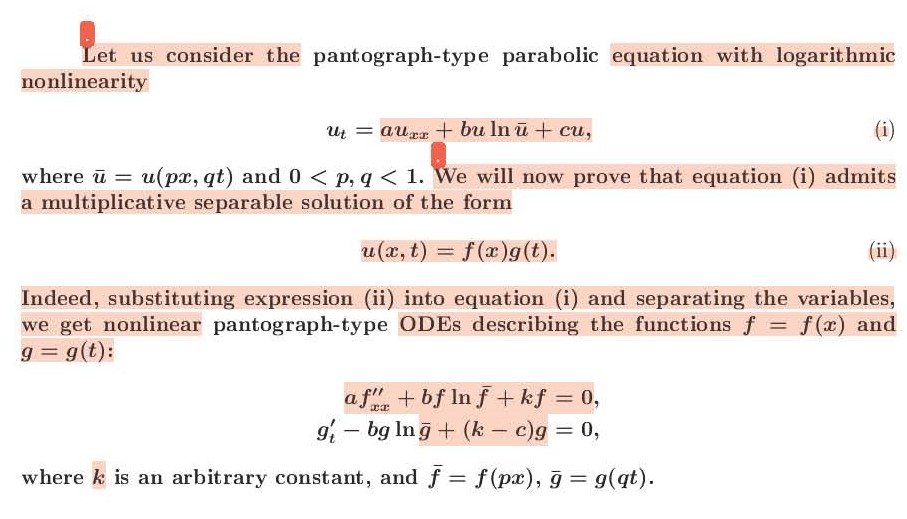}}
\label{fig:Fig4a}
\end{figure}

\medskip

\begin{figure}[!h]
\begin{flushleft}\textit{\normalsize Test problem 1, variant 2.}  \end{flushleft}
\centering
{\includegraphics[width=140mm]{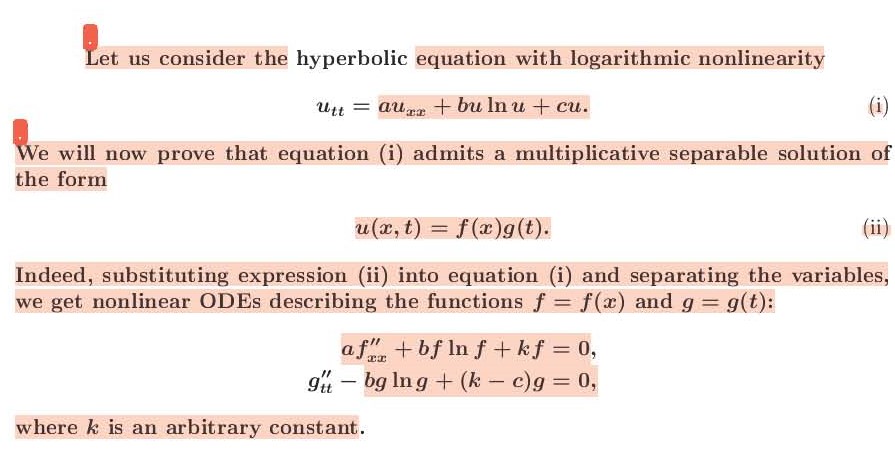}}
\label{fig:Fig4b}
\end{figure}

The iTh-system, comparing the text and equations in the above two variants
of Test problem~1, shows the following similarity indices:

69\%, if the calculation is based on counting the fragments of the text
of the 1st variant, which coincide with the fragments of the text of the
2nd variant;

93\%, if the calculation is based on counting the fragments of the text
of the 2nd variant, which coincides with the fragments of the text of the
1st variant.

These colored pictures and high similarity indices will make a strong impression
on a non-expert and he will make an erroneous conclusion: the examined texts differ
very slightly, which is convincing evidence of the existence of self-plagiarism.

\textit{Our comments}. The original equations (i) that are compared,
belong to different types and are entirely different: the equation from variant~1 is
nonlinear functional parabolic partial differential
equation with two arbitrary pantograph type delays in both independent variables
(for the first time examples of exact solutions of such equations were obtained only
in 2021 \cite{polsor2021}),
and the equation from variant~2 is a nonlinear hyperbolic partial differential equation.
The functions $f$ and $g$ determining the exact solutions of these equations with
multiplicative separation of variables also satisfy equations of entirely
different types: in variant~1, these are nonlinear pantograph type
functional ordinary differential equations of the second and first orders,
and in variant~2, these are second-order nonlinear ordinary differential equations.
The matching formulas (ii) should be excluded from the comparison at all
since they are the definition of the term ``multiplicative separable solution" \cite{polzai2012}.
In short, the content of these texts is entirely different from each other.
The considered example clearly demonstrates all the wildness and absurdity of using
the iTh-system to determine the similarity index of scientific articles
with equations and formulas.

It is important to note that the connecting words between equations and formulas
in Test problem~1 practically do not play a role.
A scientist who has written several articles on exact solutions of nonlinear
differential equations will understand the contents of both versions of this example
if the English text is replaced by German, French or Spanish,
and even if all the words are thrown away.
The same applies to any other scientific articles containing many equations and formulas:
a qualified specialist on the topic of publication usually understands
the content of the article, even if you throw out the vast majority of words from it,
but leave the equations and formulas (just like many qualified chess players
can play blindly without a chessboard).

\textit{Remark 2.}
A more detailed study of the results of the analysis of Test problem~1 additionally revealed
another important disadvantage of the iTh-system.
Namely, the obtained similarity indices significantly depend on the choice of the font
of the compared inhomogeneous texts
(above were given the similarity indices for texts in bold italics;
if you use regular italics, the corresponding similarity indices will change
and will be equal to 77\% and 94\%, respectively).

\textbf{\textit{Test problem 2.}}
To further explore the capabilities of the iTh-system,
the authors of this article came up with a multicomponent test problem
consisting of two different sets of equations and formulas.
Each set contains 30 equations and formulas, and those that are located opposite each other,
in some sense, look quite similar in appearance.
The results of the comparison of these two sets of equations and formulas are presented below.


The parts of the formulas that the iTh-system considers to be identical
to the corresponding parts of the formulas in the left column
are colored in red in the right column.
The final result that the iTh-system gives is simply stunning:
the inhomogeneous text in the right column is 87\% the same as the inhomogeneous text in the left column
(the calculation is based on counting the text fragments of the
right column that coincide with the text fragments of the left column).
It can be seen that the iTh-system in fifteen cases was unable
to distinguish between different equations and formulas
(which is 50\% of the total tested set of formulas).
Similarly, it was found that the inhomogeneous text
in the left column coincides by 71\% with the text in the right column.

Thus, the iTh-system recommends the user to make an erroneous conclusion
that the equations and formulas in both sets of the considered test problem
differ little. Since all formulas and equations in both sets are different,
we once again come to the obvious conclusion: the iTh-system cannot be used
to compare texts that contain many equations and formulas.

\medskip
\begin{figure}[!htbp]
\begin{flushleft}\textit{\normalsize Test problem 2.}  \end{flushleft}
\centering
{\includegraphics[scale=0.7]{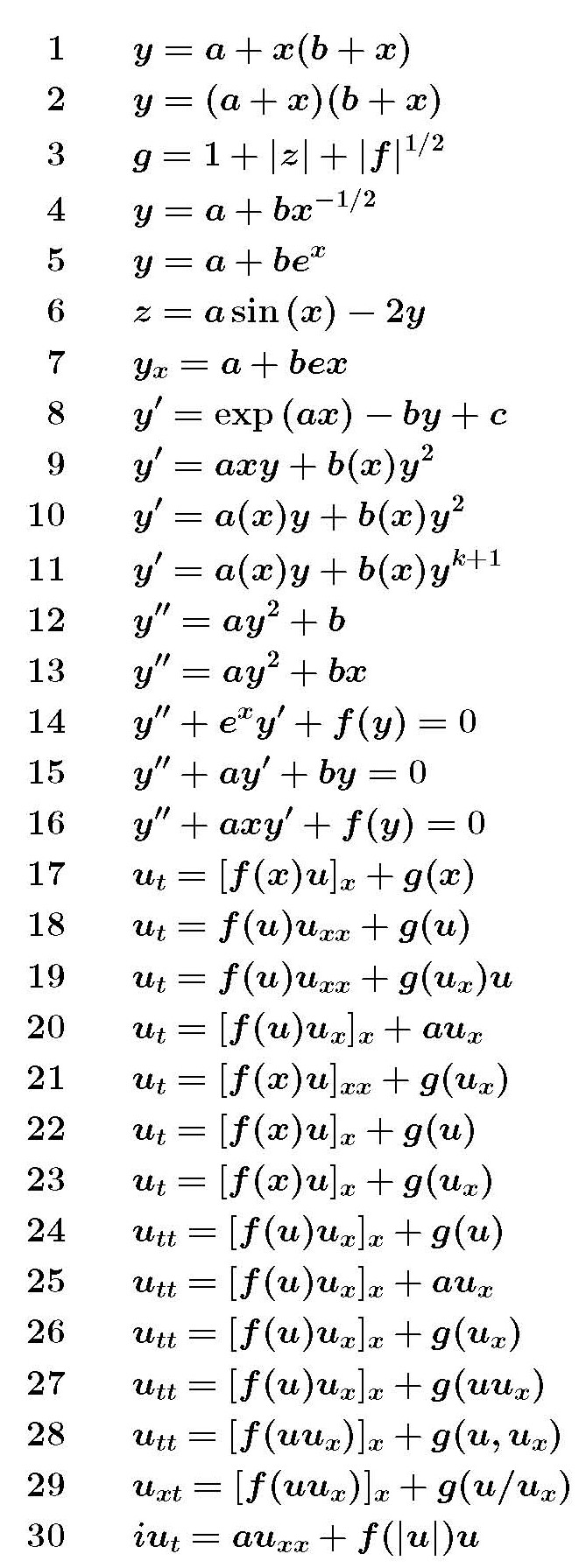}\quad \includegraphics[scale=0.825]{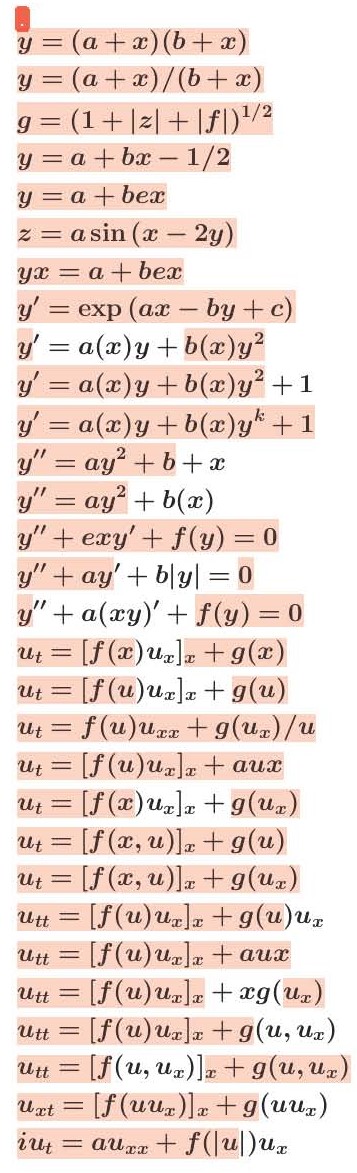}}
\label{fig:FigK}
\end{figure}

\textit{Remark 3.} In Test problem~2, as in Test problem~1,
the similarity indices obtained by using the iTh-system significantly depend
on the choice of the font of the matched inhomogeneous texts
(the similarity indices for equations typed in bold italics were given above;
if you use regular italics, then the corresponding indices will be 78\% and 75\%).

While working on this part of the article, one of the authors had a bad dream.
Namely, the exam on the topic \textit{Equations of mathematical physics} is coming to an end.
The student responds poorly, and the professor (an author of this article),
feeling sorry for the student and trying to give him a satisfactory grade, asks him to write
the wave equation (i.e. $u_{tt}=au_{xx}$).
The student, after some thought, writes the heat equation $u_{t}=au_{xx}$.
Naturally, the professor gives the student an unsatisfactory grade.
An hour later, the professor summoned to the dean's office
to discuss the student's complaint, in which he wrote:
``I just missed one small subscript $t$ in the equation.
According to the iTh-system, the similarity index of
the equation I wrote is as much as 87.5\%, which is a very good result" \ldots

Imagine now that the absolutely absurd ideology of comparing
individual fragments (as the iTh-system does with formulas)
will be transferred to artists and their paintings.
Then, for example, the world-famous Russian marine artist Ivan K. Aivazovsky
will be declared a great plagiarist and self-plagiarist;
indeed, almost all of his paintings
depict water, waves, clouds, and, sometimes, ships
(in a similar way, the famous English romantic landscape artist Joseph M. W. Turner and
the first Norwegian great romantic landscape artist J. Christian C. Dahl can also be attributed to self-plagiarists).
And the poor portrait painters: after all, they all paint only the forehead,
cheeks, mouth, eyes, ears, and hair (sometimes clothes).

It is curious to note that the developers of the iTh-system
compare formulas by fragments, but words do not.
The question is, why did such an unfair discrimination of mathematical formulas take place?

Let us demonstrate what will happen if words to compared by fragments.

\textbf{\textit{Example 4.}}  Let us take two phrases that are completely different in meaning:

``\red{The} \red{solut}e \red{was} in a con\red{taine}r" and ``\red{The} exact \red{solut}ion \red{was} ob\red{taine}d".

There are 24 letters in the first phrase,
16 of them in fragments (they highlighted in color)
coincide with the letters of the second phrase.
Therefore, the similarity index of the first phrase (when compared with the second phrase) is more than 66\%.

What kind of nonsense is this, you say?
But this is precisely how the iTh-system works with mathematical equations and formulas!

\subsection{Difficult situations requiring the involvement of highly qualified specialists}

In many cases, it is only a highly qualified specialist
who is well versed in the topic (which can be quite narrow) of the article in question
to determine correctly whether a given coefficient of the equation
is insignificant or very important.
Moreover, adequate conclusions may differ depending on the field of research conducted.
Let us illustrate the above with a simple concrete example.

\textbf{\textit{Example 5.}}
Let us consider the Abel differential equation of the second kind with quadratic nonlinearity
\begin{align}
yy'_{x}-y=ax+bx^2,
\label{eq:04}
\end{align}
where $a$ and $b$ are free parameters. Two qualitatively different situations are possible.

1. If the Cauchy problem for equation \eqref{eq:04} is numerically solved,
then the specific values of the parameters $a$ and $b$ are insignificant.
In this case, two equations of the form \eqref{eq:04} for different values of
the parameters $a$ and $b$ can be considered similar.

2. If questions of the integrability of equation \eqref{eq:04} are considered,
then the values of the parameter $b$ are inessential, and the values of $a$ are essential.
Currently, only two values $a=\pm \frac{6}{25}$ are known,
for which equation \eqref{eq:04} admits a closed-form solution~\cite{polzai2018}.
Therefore, if in some paper the integrability of this equation
for other values of $a$ is proved, this result will certainly be new.
It is obvious that the iTh-system will draw wrong conclusions in this case.

This example illustrates well the cardinal qualitative differences
in publications devoted to numerical and exact solutions of mathematical equations.
Namely, the specific values of the coefficients of the equations
under consideration are usually of little importance when using numerical methods,
but, as a rule, they are very important when using exact analytical methods.

\textit{Remark 4}. Using handbooks \cite{polzai2018,polzai2012}, which contain many
exact solutions of ordinary and partial differential equations,
it is not difficult to give other examples of nonlinear equations that have qualitative features similar to Example~5.

The correct interpretation of the results of processing articles
(on integrability and exact solutions of ordinary differential equations
or partial differential equations) by using the iTh-system
can only be given by a highly qualified specialist specializing in this field
(to analyze the correctness or incorrectness of the results of processing
such publications by the iTh-system, for example, should not involve
mathematicians who are specialists in numerical methods,
differential geometry, and number theory).
And, of course, neither technical assistants of the editor, nor
administrative staff who do not have special knowledge,
unable to do this.

\section{The iThenticate system does not take into account plots, figures, and drawings}

The iTh-system does not take into account plots, figures, and drawings
(they are simply discarded), which is completely wrong.
Plots, figures, and drawings are often more important and descriptive than
the verbal text of the article describing them.
The error of ignoring drawings is illustrated by a simple example.

\textbf{\textit{Example 6.}}
In the elementary education book for children, there are two drawings:
the first depicts a cat, and the second a dog (see below).
The text under both drawings is the same: ``The picture shows a pet
that has four legs, two eyes, two ears, nose, mouth, tail,
is covered with wool and eats meat. Write the name of this animal in your notebook."
Since the iTh-system does not take drawings into account,
it will come to the ridiculous conclusion that there is 100\% similarity
in these texts with different drawings (i.e. cat = dog).
Obviously, in the considered example, the images are much more
important than the related text.

\medskip
\begin{figure}[!htb]
\centering
{\includegraphics[height=60mm]{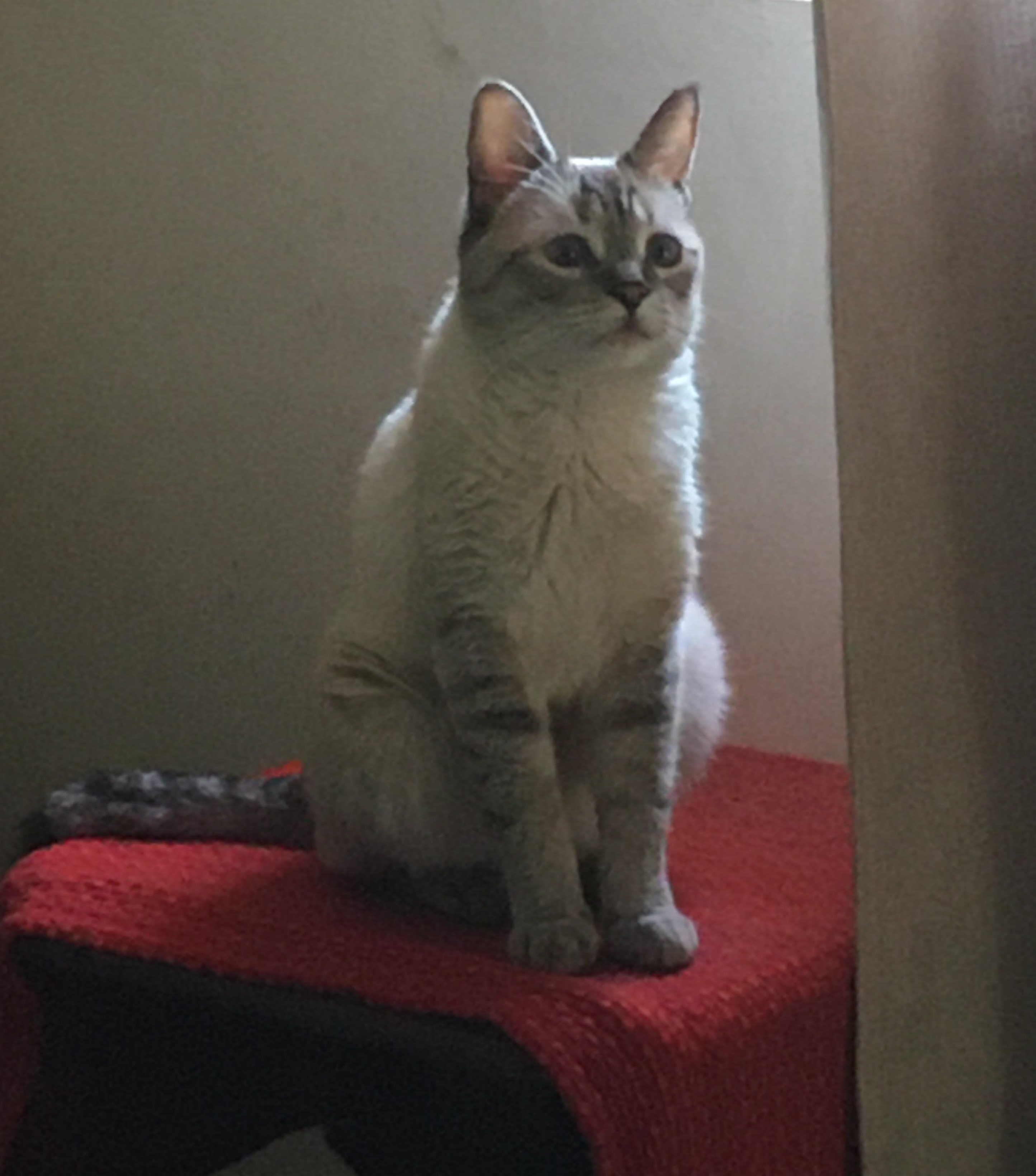}\quad \includegraphics[height=60mm]{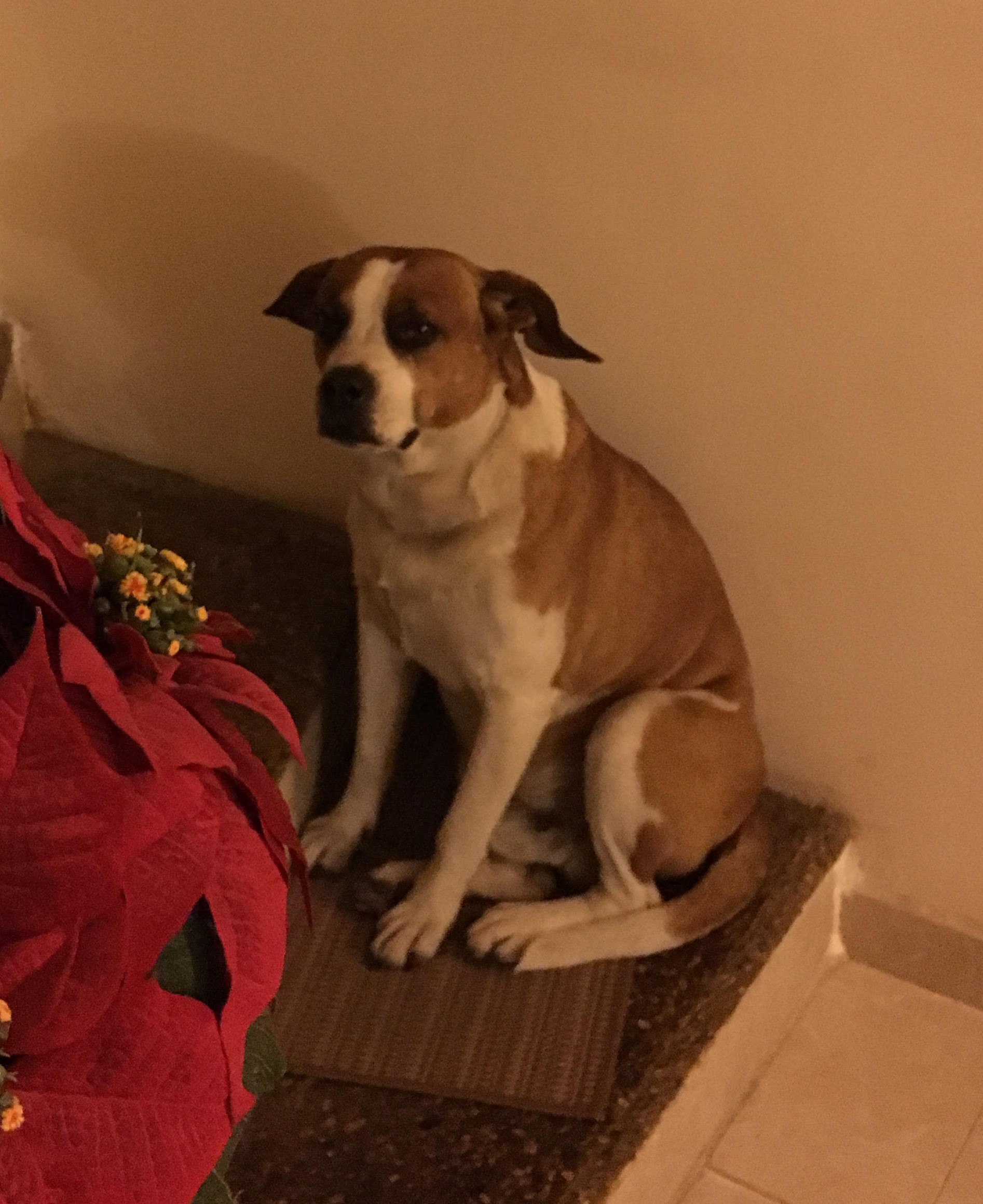}}
\label{fig:FigN}
\end{figure}

It is important to note that sometimes plots and drawings
can form the main content of an article or be an important integral
part of it\footnote{This is especially true for experimental work,
as well as publications devoted to the numerical simulation
of specific structures and devices.}.
Therefore, they must be taken into account when
evaluating the similarity index of an article.
Technically, this is not difficult to do, proceeding, for example,
from the area occupied by plots and drawings and the area of the text of an article
processed by the iTh-system without plots and drawings.

Please note that since the iTh-system cannot compare drawings and images, therefore it cannot also identify possible copyright violations.

When mathematical tables are processed by the iTh-system,
the equations and formulas contained in them, are compared by individual
fragments and letters, just like equations and formulas are compared
in the text of the article (the disadvantages of such a comparison
are described in detail in the previous section).

\section{Possible methods to compare inhomogeneous texts with equations and formulas}\label{ss:ways-to-improve}

For simplicity and clarity, we will restrict ourselves to
considering inhomogeneous texts with a large number of equations and formulas,
but without figures, drawings, and tables. We also assume that the software
system used is capable of identifying identical equations and formulas in different texts.

Two possible methods of comparing inhomogeneous texts with equations and formulas
by using software systems such as the iTh-system are described below.

\textit{Method 1.}  It is necessary to discard the entire verbal text
from the article under consideration and compare the remaining equations and formulas
with equations and formulas in other articles. In this case,
it is necessary to use the basic principle of comparison:

\textit{Two equations or formulas are considered the same if all letters, numbers, and
mathematical symbols} included in them are the same.
(\textit{Any equation or formula is a single whole, and it cannot
be compared by breaking them into constituent fragments and letters.})

This simple method has one very important advantage over any others
since it allows one to compare inhomogeneous texts in different
languages (since letters, special mathematical symbols, and numbers in equations and formulas
do not change).
It should be noted that, until now, such texts have not lent
themselves to comparison with the help of existing software systems.

\textit{Method 2.}
You must first specify how many words one formula is equivalent to and then compare the inhomogeneous text
using the principle of comparing equations and formulas formulated in Method~1.
We believe that one formula should cost at least 5--10 times more (and possibly even more) then one word.

It is best to provide for the possibility of using any of the two
above-described methods for comparing inhomogeneous texts
by the software system at the request of the user.

The implementation of the above-described possible methods for comparing
inhomogeneous texts will significantly improve the operation of software
systems such as the iTh-system with scientific publications
containing many equations and formulas.

\section{Final conclusions and some remarks}

The logical reasoning and specific examples presented in this article,
and the analysis of the results of processing our test problems by the iTh-system
allow us to conclude that this system is very ineffective for assessing
the similarity index of inhomogeneous scientific articles containing
equations, formulas, figures, drawings, and tables.
In this regard, the following should be noted:

1. You cannot blindly trust the results of applying the iTh-system to
scientific articles with inhomogeneous text, since the color highlighting
of equations and formulas can be erroneous and must be checked.
The extensive research carried out in this article gives all the grounds
to argue that this software system can overestimate the similarity index
of articles with the inhomogeneous text by several times.

2. Any estimates of the similarity index of scientific articles containing
a significant number of equations, formulas, figures, and drawings,
based on the use of the iTh-system (and any similar software systems,
both existing and those that will appear in the future)
should be very carefully checked by highly qualified specialists
on the topic of the articles under consideration.

3. Full results of processing an article with formulas by the iTh-system
(in the case of a high similarity index) must be sent to the authors
without fail, so that they have the opportunity to check these results
for adequacy and reasonably challenge them.

\textit{Remark 5.} The iTh-system is effective only for detecting the obvious
self-plagiarism of inhomogeneous text, when the author includes in his article
meaningful in content and significant in volume pieces of text from his other publications
without the necessary reference to them.

In other words, the iTh-system in no way solves the problem of identifying
self-plagiarism of authors of scientific articles with an inhomogeneous text
containing a significant number of equations and formulas.
Moreover, the use of this system in practice has created
an unnecessary time-consuming problem of scrupulous manual check of the adequacy of its work.

A characteristic distinguishing feature of the development of modern science is that
often scientists cannot fully evaluate the results of colleagues
working in seemingly very close areas
(for example, specialists in differential equations, as a rule,
don't understand too much in integral and functional equations,
moreover, specialists in partial differential equations
usually cannot be used as reviewers for
ordinary differential equations papers, and vice versa).
Therefore, highly qualified reviewers cannot be replaced by technical workers
and administrative-scientific managers,
even if they are provided with a text comparison software,
such as the iTh-system.

It is important to note that the unfair selective application
of the iTh-system to the texts of articles of some (but not all) authors
can serve as a tool for discriminating authors on gender, racial, ethnic, and other grounds.

\section*{Acknowledgments}


The authors are grateful to A.V. Aksenov, A.L. Levitin, and A.N. Filippov
for their attention to the work and useful discussions.

\end{document}